# Synthesizing Correlations with Computational Likelihood Approach: Vitamin C Data

Myung Soon Song

*Department of Mathematics, SUNY Cortland, Cortland NY, USA*

myungsoon.song@cortland.edu, 607-753-4569

**0. Abstract**

It is known that the primary source of dietary vitamin C is fruit and vegetables and the plasma level of vitamin C has been considered a good surrogate biomarker of vitamin C intake by fruit and vegetable consumption. To combine the information about association between vitamin C intake and the plasma level of vitamin C, numerical approximation methods for likelihood function of correlation coefficient are studied. The least squares approach is used to estimate a log-likelihood function by a function from a space of *B*-splines having desirable mathematical properties. The likelihood interval from the Highest Likelihood Regions (HLR) is used for further inference. This approach can be easily extended to the realm of meta-analysis involving sample correlations from different studies by use of an approximated combined likelihood function. The sample correlations between vitamin C intake and serum level of vitamin C from many studies are used to illustrate application of this approach.

# 1. Introduction

How to combine information about population correlations from many different independent studies is one of the hot questions in meta-analysis. Conventionally, asymptotic methods are used to tackle this problem. But these approaches are naïve and have some evident flaws. One of the most serious of these flaws is that we must assume that the sample size from each study is sufficiently large to justify Central Limit approximations, an assumption which is violated in many situations. [12]

In this paper, approximation methods for calculating the combined likelihood of correlations from different studies is introduced with the vitamin C data.

Recently, a great deal of attention has been given to fruit and vegetable consumption and their role in reducing rates of chronic diseases such as cancer, coronary heart disease (CHD), stroke, diabetes, and arthritis. It is suggested that the protective effect of fruits and vegetables is partly due to antioxidant nutrients like vitamin C and carotenoids which inhibit lipid per-oxidation and oxidative cell damage. [14]

Vitamin C intake is measured by many different methods including Food Frequency Questionnaires (FFQ), Diet History Questionnaires (DHQ), 24 hr Dietary Recalls (DR) and Weight Records (WR). FFQ is one of the most commonly used tools in epidemiologic studies to assess long-term nutritional exposure. It is used to determine usual intakes of selected items from all major food groups.

Because fruits and vegetables are the main source of dietary vitamin C, the serum level of vitamin C has been considered a good predictor of vitamin C intake from fruit and vegetable consumption. Therefore, significant changes of plasma vitamin C are expected by altering fruit and vegetable consumption.

Dehghan et al. [3] assess the association between vitamin C intake measured by dietary assessment methods and plasma level of vitamin C in epidemiological studies. The purposes of their study are: first, to investigate the strength of the correlation between plasma level of vitamin C as a biomarker and dietary vitamin C intake measured by methods commonly used for dietary assessment in epidemiological studies and, second, to explore whether the correlation between dietary vitamin C intake and plasma vitamin C varies between different dietary assessment methods.

In the following sections, a meta-analysis will be given based on the data from Dehghan et al. [3]. Table 1 shows the details about the data sets. The analysis is done for the ten studies of the correlation between dietary vitamin C measured and plasma vitamin C for both of males and females.

A *B*-spline approximation method for developing a parsimonious model of the correlation structure is suggested.

| Table 1: Ten Studies used for Vitamin C example | | | |
|---|---|---|---|
| Study ID | Sample | $r_{(i)}$ | $n_{(i)}$ |
| 1 | Boeing et al. (1997) | 0.24 | 92 |
| 2 | Cooney et al. (1995) | 0.25 | 27 |
| 3 | Drewnowski et al. (1997) | 0.29 | 837 |
| 4 | EPIC group of Spain (1997) | 0.46 | 72 |
| 5 | Jacques et al. (1993) | 0.39 | 139 |
| 6 | Katsouyanni et al. FFQ1 (1997) | 0.30 | 80 |
| 7 | Malekshah et al. (2006) | 0.35 | 131 |
| 8 | McKeowen et al. FFQ1 (2001) | 0.42 | 118 |
| 9 | McKeowen et al. FFQ2 (2001) | 0.39 | 118 |
| 10 | Schroder et al. (2001) | 0.53 | 44 |

## 2. Explanation of some tools used

In this section, some tools used in the following sections are explained briefly. If you know these already, you can skip this section.

*2.1. B-spline*

In mathematics or statistics, a *B*-spline is a spline function having minimal support with respect to a given degree, smoothness, and domain partition.

The term *B*-spline was first introduced by Romanian mathematician Isaac Jacob Schoenberg and is short for basis spline. *B*-splines can be evaluated in a numerically stable way by the de Boor algorithm [2].

Hastie et al. [6] provides a quite rigorous definition of the *B*-spline. Let $\xi_0$ and $\xi_{K+1}$ be two boundary knots, which typically define the domain over which a spline is evaluated. Now the augmented knot sequence $\tau$ satisfies:

1. $\xi_0 < \xi_1$ and $\xi_K < \xi_{K+1}$
2. $\tau_1 \leq \tau_2 \leq \cdots \leq \tau_M \leq \xi_0$
3. $\tau_{j+M} = \xi_j, \quad j = 1, \cdots, K$

4. $\xi_{K+1} \leq \tau_{K+M+1} \leq \tau_{K+M+2} \leq \cdots \leq \tau_{K+2M}$.

The actual values of these additional knots beyond the boundary are arbitrary, and it is customary to make them all the same and equal to $\xi_0$ and $\xi_{K+1}$ respectively.

Denote by $B_{i,m}(x)$ the $i^{th}$ B-splines basis function of order $m \leq M$ for the knot-sequence $\tau$. They are defined recursively in terms of divided differences as follows:

$$B_{i,1}(x) = \begin{cases} 1 & \text{if } \tau_i \leq x \leq \tau_{i+1} \\ 0 & \text{otherwise} \end{cases} \quad (1)$$

for $i = 1, \cdots, K+2M-1$.

$$B_{i,m}(x) = \frac{x - \tau_i}{\tau_{i+m-1} - \tau_i} B_{i,m-1}(x) + \frac{\tau_{i+1} - x}{\tau_{i+m} - \tau_{i+1}} B_{i+1,m-1}(x) \quad (2)$$

for $i = 1, \cdots, K+2M-m$.

Thus, with $M = 4$, $B_{i,m}$, $i=1, \cdots, K+4$ are the $K+4$ cubic B-spline basis functions for the knot sequence $\xi$. This recursion can be continued and will generate the B-spline basis for any order spline. In the statistical package R, cubic spline basis is used as default for B-spline approximation.

A fundamental theorem states that every spline function of a given degree, smoothness, and domain partition can be represented as a linear combination of B-splines of that same degree and smoothness, and over that same partition.

Also, a B-spline curve of order $K$ is in general $C^{K-2}$ continuous (continuous up to $(K-2)^{th}$ derivative). For example, a cubic B-spline curve (a B-spline curve of order 4) is $C^2$ continuous. At a knot position the continuity is $C^{K-M-1}$ where $M$ is the multiplicity of that knot.

In this paper, the cubic B-spline is used for approximation. From now on, a B-spline refers to a cubic B-spline if there is no further explanation.

## 2.2. Highest Likelihood Regions (HLR)

Many statistical methods involve summarizing a probability distribution by a region of the sample space covering a specified probability. But it is not always clear which region should be used. Suppose someone wishes to give a *90%* prediction interval from a given distribution. Should he or she use the interval symmetric with respect to the mean or the median, the interval defined between the *5%* and *95%* quantiles, the interval of shortest length, or the interval that maximizes the probability of covering a given set? Hyndman [10] investigates an approach to this question by suggesting highest density regions (HDR) from any given (possibly multivariate) density *f(x)* which is bounded and continuous in x. He claims that the usual purpose in summarizing a probability distribution by a region of the sample space is to sketch a comparatively small set which contains most of the specified probability. The criteria he adapts are the following:

1. The region should occupy the smallest possible volume in the sample space.

2. Every point inside the region should have probability density at least as large as every point outside the region.

It follows immediately from the criteria and corresponding definition of HDR that the highest density region has the smallest possible volume in the sample space. Furthermore, the HDR has some virtues like:

1. The modes are contained in every HDR.
2. The HDR has a link to conventional wisdom. For example, in the case of a normal distribution an HDR coincides with the usual probability region symmetric with respect to the mean spanning the $\frac{\alpha}{2}$ and $1-\frac{\alpha}{2}$ quantiles. This assertion is true for any symmetric unimodal distribution.

In this paper, I will use the term Highest Likelihood Regions (HLR) rather than the Highest Density Regions (HDR). Both density functions and likelihood functions are non-negative on the corresponding domains. But there is a big difference between these two types of function. The area under the density curve defined on the domain should equal one, but the likelihood function has no such restriction. Of course, the essential characteristics of HLR are inherited from those of HDR with no difficulty.

## 3. Background

In many areas of research, it is useful to assess the relationship between continuous variables. Correlation coefficients have been used extensively as an index of the relationship between two normally distributed variables. Since the correlation coefficient is a scale-free measure of the relationship between variables, it is invariant under substitution of different but linearly equitable measures of the same construct. Therefore, the correlation coefficient is a natural candidate as an index of effect magnitude across studies in meta-analysis. [7]

How to combine the sample correlation coefficients from many independent studies having possibly different sample sizes has been an old question in meta-analysis.

Arguably, the most common process to deal with the issue above in meta-analysis is (P1) to calculate the sample correlations for individual studies, (P2) convert them to a common metric, and (P3) combine the results to obtain an average effect size. Why the process (P2) is needed instead of going straight from (P1) to (P3) may be worthy of explanation.

The sample correlation $r$ was proposed by Pearson as an estimator of the population correlation $\rho$. The exact distribution of the sample correlation coefficient under the assumption of a bivariate normal distribution was first derived by Fisher [4], who obtained the distribution in a rather complicated form. Simpler forms of the distribution more suitable for computation are given by Hotelling [8]. These distributions will be shown later in this section. Due to the complexity of the exact distribution, the large sample distribution of a sample correlation $r$ has been preferred. The asymptotic distribution of a sample correlation $r$ is normal with mean $\rho$ and variance, $\frac{(1-\rho^2)^2}{n}$ where $n$ is the sample size. Unfortunately, the variance of $r$ in the approximation depends strongly on $\rho$, the unknown true value of the correlation. To stabilize the variation of $r$, Fisher [5] proposes the $z$-transformation.

$$z = z(r) = \frac{1}{2}\log\frac{1+r}{1-r} = \tanh^{-1} r \qquad (3)$$

The corresponding transformation for $\rho$ is

$$\zeta = \zeta(\rho) = \frac{1}{2}\log\frac{1+\rho}{1-\rho} = \tanh^{-1}\rho \qquad (4)$$

The $z$-transformation stabilizes the variance in the sense that $z$ is approximately normally distributed with mean $\zeta$ and variance $\frac{1}{n}$ when $n$ is large. A more accurate approximation to the distribution of $z$ is obtained by setting the asymptotic variance equal to $\frac{1}{n-3}$ instead of $\frac{1}{n}$ for moderate values of $n$. Consequently, $\sqrt{n-3}(z-\zeta)$ has approximately the standard normal distribution:

$$\sqrt{n-3}(z-\zeta) \sim N(0,1) \qquad (5)$$

For the conversion of metric (P2) described in the previous page, there are some alternative approaches (such as unbiased estimators, Kramer's *t*-transformation) beside Fisher's *z*-transformation. [7] Among these, Fisher's *z*-transformation is the most popular method, partly because of a simple distributional form. But Fisher's *z*-transformation produces an upward bias in the estimation of the correlation coefficients in the process (P3) described in the previous page. This upward bias is usually higher than the negligible downward bias produced by untransformed correlations. [9] Furthermore, Fisher's *z*-transformation is based on asymptotic theory, and may not work very well for small sample sizes.

Considering these problems, asking whether direct approaches from the process (P1) to the process (P3) are possible is still attractive. In the following sections, the use of log-likelihood functions and the *B*-spline approximations are suggested to answer this question.

### 4. Approach – Likelihood Function and Approximation

It is well known that likelihood functions play an important role in both the frequentist and Bayesian statistical paradigms. Many times, uncertainty is considered by using the likelihood when studying a statistical problem. The concept of the likelihood is one of the best methods for unifying the demands of statistical modeling and inference. One of the very important advantages of likelihood functions is that they are most naturally represented, understood, and communicated graphically. To see what the data say, we look at graphs of likelihood functions. [11]

The distribution of the sample correlation *r* in a sample of size *n* from a bivariate normal distribution with correlation $\rho$ was first obtained by Fisher [4] in the form:

$$\frac{(1-\rho^2)^{\frac{1}{2}(n-1)}(1-r^2)^{\frac{1}{2}(n-4)}}{\pi(n-3)!} \left| \frac{d^{n-2}}{dx^{n-2}} \left\{ \frac{\cos^{-1}(-x)}{\sqrt{1-x^2}} \right\} \right|_{x=r\rho} \tag{6}$$

Anderson [1] gives a different form of the density,

$$\frac{2^{n-3}(1-\rho^2)^{\frac{1}{2}(n-1)}(1-r^2)^{\frac{1}{2}(n-4)}}{\pi(n-3)!} \sum_{\alpha=0}^{\infty} \frac{(2\rho r)^\alpha}{\alpha!} \Gamma^2\left[\frac{1}{2}(n-1+\alpha)\right] \tag{7}$$

but the most commonly used form of the likelihood function of the population correlation coefficient based on a single sample correlation *r* from a sample size *n* is

$$L(\rho \mid r, n) = \frac{n-2}{\sqrt{2\pi}} \frac{\Gamma(n-1)}{\Gamma\left(n-\frac{1}{2}\right)} (1-\rho^2)^{\frac{1}{2}(n-1)} (1-r^2)^{\frac{1}{2}(n-4)} (1-\rho r)^{-n+\frac{1}{2}}$$
$$\times F\left(\frac{1}{2}, \frac{1}{2}; n-\frac{1}{2}; \frac{1+\rho r}{2}\right) \quad (8)$$

where

$$F(a,b;c;x) = \sum_{a=0}^{\infty} \frac{\Gamma(a+j)}{\Gamma(a)} \frac{\Gamma(b+j)}{\Gamma(b)} \frac{\Gamma(c)}{\Gamma(c+j)} \frac{x^j}{j!} \quad (9)$$

which stems from Hotelling [8] who made a comprehensive study of the distribution of $r$. The series in Eqn (9) converges faster than the ones in (6) or (7).

Fig 1 displays the likelihood functions for $\rho$ when the sample correlations $r$ range from 0.1 to 0.9 in steps of 0.1, respectively with a sample size $n=100$. The greater the magnitude of the sample correlation, the more the likelihood function concentrates around the peak or mode.

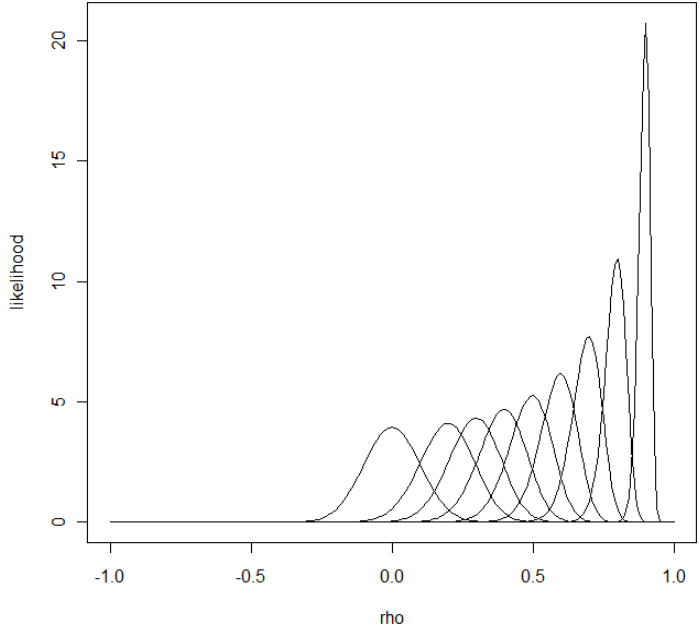

*Figure 1. Likelihood functions obtained from positive sample correlations*

Fig 2 displays the log-likelihood functions of $\rho$ and their $B$-spline approximations for fixed sample size $n=100$ and $r=0$ when one knot is used. The solid line is for the log likelihood function and the dashed line is for the $B$-spline approximation.

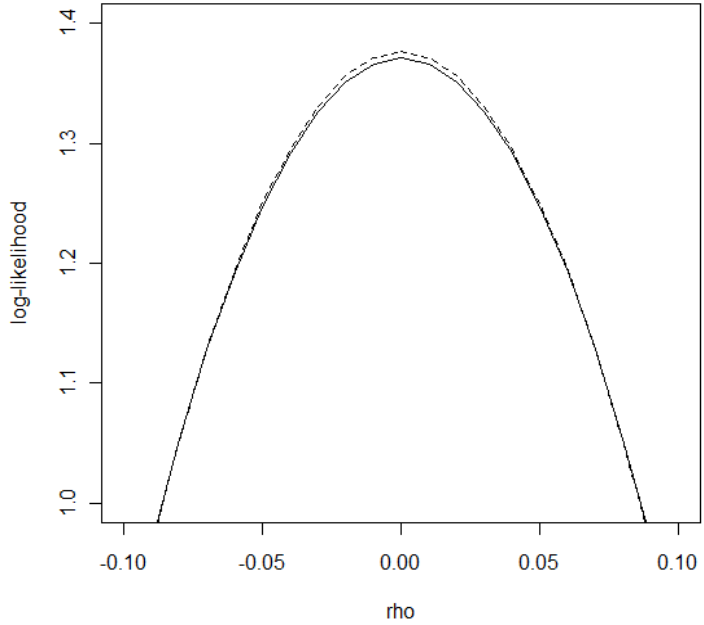

*Figure 2. The log-likelihood (solid one) and its B-spline approximation (dashed one)*

At this point, we need to mention why the *B*-spline approximation is needed. It will be ideal if we can find the MLE (or can find them by taking the first derivative and the second derivative from the likelihood function). But as you can see in (8), it is extremely hard to get the derivatives because the likelihood has a very complicated mathematical form. It seems that numerical methods are plausible to calculate the sum of the infinite series.

If numerical methods are used, it is natural to ask how to truncate the number of terms in the series or how to control the error in computing the likelihood. If we rewrite the infinite series representation of the likelihood in (8) as:

$$\sum_{m=0}^{\infty} a_m = \sum_{m=0}^{n} a_m + \sum_{m=n+1}^{\infty} a_m \qquad (10)$$

where $a_m$ is the $m^{th}$ term of the infinite series representation in (8).

then the remainder (or error) $\sum_{m=n+1}^{\infty} a_m$ satisfies the inequalities:

$$\int_{n+1}^{\infty} a_m dm \leq \sum_{m=n+1}^{\infty} a_m \leq a_{m+1} + \int_{n+1}^{\infty} a_m dm \qquad (11)$$

by the Maclaurin-Cauchy integral test.

Now it turns out that we can control the size of the remainder by controlling the size of the integration in the inequalities. This goal can be achieved by using iteration algorithms and quadrature approximation methods suggested in Song [12] and Song and Gleser [13].

## 5. Application – Vitamin C

In this section $X$ and $Y$ will be random variables with bivariate normal distribution. Consider the situation in which each of $k$ studies has examined correlations between $X$ and $Y$. Let $r_{(i)}$ and $n_{(i)}$ be the sample correlation coefficient between $X$ and $Y$ and the sample size from the $i^{th}$ study for $i = 1, \cdots, K$, respectively.

### 5.1. Homogeneity Test

If $k$ studies investigate the same population correlation, combining the data from several studies to form a single estimated correlation is meaningful. Thus, we first need to determine whether the data obtained from several studies are consistent with the hypothesis of homogeneous correlations.

A formal hypothesis test can be made for the hypotheses:

$$\begin{cases} H_0: & \rho_1 = \rho_2 = \cdots = \rho_9 = \rho_{10} \\ H_a: & \text{Not all of } \rho_i\text{'s are the same} \end{cases} \quad (11)$$

where $\rho_i$ is a population correlation from the $i^{th}$ study for $i = 1, \cdots, 10$ This test uses the statistic

$$Q = -2\log \left[ \frac{\sup_{H_0} \mathbf{L}(\boldsymbol{\rho} \mid r)}{\sup_{H_a} \mathbf{L}(\boldsymbol{\rho} \mid r)} \right] \quad (12)$$

where $L(\boldsymbol{\rho} \mid \mathbf{r}) = \prod_{i=1}^{10} L(\rho_i \mid r_i) = \exp\left[\sum_{i=1}^{10} \log L(\rho_i \mid r_i)\right]$ is the combined likelihood from the ten studies.

$Q$ has approximately a $\chi^2$ distribution with 9 $(=10-1)$ degrees of freedom if $H_0$ is true. A test of $H_0$ at the $100\alpha$ % level of significance is given by rejecting $H_0$ if $Q$ is greater than the $100(1 - \alpha)$ percentile of the $\chi^2$ distribution with 9 degrees of freedom.

By using the statistical software R, we obtain $Q = 9.40$ which is less than 16.92, the 95% critical value of the $\chi^2$ distribution with 9 degrees of freedom, so there is no enough evidence to reject $H_0$ in favor of $H_a$ at the 5% level of significance.

*5.2. Inference and Comparison of Estimators*

In Table 2, 95% confidence intervals by using conventional asymptotic theory, 95% HLR likelihood intervals of the exact likelihood function and 95% HLR likelihood intervals obtained from the B-spline approximation to the log-likelihood function are compared, for each study. Table 2 shows differences between the classical asymptotic method and the two likelihood methods. Generally, the intervals from the asymptotic method tend to shift to the right compared to those of the likelihood methods for all studies. As we see, B-splines approximate the corresponding exact likelihood function very well.

When a series of k independent studies share a common population correlation coefficient $\rho$, the method usually used to estimate the $\rho$ is to transform each r by a z-transform to yield $z_1, \cdots, z_k$ given in (3) and then calculate the weighted average:

$$z_w = w_1 z_1 + w_2 z_2 + \cdots + w_k z_k \qquad (13)$$

where $w_i = \dfrac{(n_i - 3)}{\sum_{m=1}^{k}(n_m - 3)}$ . [7]

| Table 2: 95% C.I's and HLR likelihood intervals from individual studies for comparison ||||
|---|---|---|---|
| Study ID | Asymptotic | Exact Likelihood | B-spline Approximation |
| 1 | (0.037, 0.424) | (0.048, 0.413) | (0.048, 0.413) |
| 2 | (0.000, 0.575) | (0.000, 0.509) | (0.000, 0.508) |
| 3 | (0.227, 0.351) | (0.227, 0.351) | (0.227, 0.350) |
| 4 | (0.256, 0.625) | (0.255, 0.621) | (0.255, 0.623) |
| 5 | (0.239, 0.523) | (0.239, 0.520) | (0.239, 0.521) |
| 6 | (0.086, 0.488) | (0.091, 0.481) | (0.091, 0.481) |
| 7 | (0.190, 0.492) | (0.190, 0.489) | (0.190, 0.490) |
| 8 | (0.259, 0.558) | (0.259, 0.556) | (0.259, 0.557) |
| 9 | (0.225, 0.533) | (0.225, 0.530) | (0.225, 0.531) |
| 10 | (0.277, 0.714) | (0.275, 0.709) | (0.275, 0.712) |

The corresponding estimate $\tilde{r}$ is given by

$$\tilde{r} = \frac{e^{z_w} - 1}{e^{z_w} + 1} \tag{14}$$

Table 3 summarizes the main results obtained from the statistical software *R* from this way of combining the ten studies. It shows the Maximum Likelihood Estimate $\hat{r}$ (MLE hereafter) from the exact combined likelihood or its *B*-spline approximation and the asymptotic pooled estimate $\tilde{r}$. Note that $\hat{r}$s are different from $\tilde{r}$ by 0.009. Also, in Table 3 appear the 95% asymptotic C.I. for $\rho$ and the 95% HLR likelihood intervals for $\rho$. The lower limit and the upper limit of the asymptotic C.I. are less than the counter parts in the HLR likelihood intervals, for each.

**Table 3: Comparing Results from Combined Study**

|  | Asymptotic | Exact Likelihood | *B*-spline Approximation |
|---|---|---|---|
| Point Estimator | $\tilde{r} = 0.332$ | $\hat{r} = 0.341$ | $\hat{r} = 0.341$ |
| Interval Estimator | (0.288, 0.374) | (0.289, 0.390) | (0.289, 0.390) |

## 6. Simulation – Vitamin C

In this section, a simple simulation is conducted to compare the accuracies of the asymptotic pooled estimate $\tilde{r}$ and the MLE $\hat{r}$ from the combined likelihood in terms of Mean Squared Error (MSE hereafter).

The MSE of $\tilde{\rho}$ with respect to $\rho$ is defined as:

$$MSE(\tilde{\rho}) = E(\tilde{\rho} - \rho)^2 \tag{15}$$

where $\rho$ is the population correlation and $\tilde{\rho}$ is the estimate of $\rho$. ($\tilde{\rho}$ can be the asymptotic pooled estimate $\tilde{r}$ or the MLE $\hat{r}$ from the combined likelihood.)

In the simulation, eight different cases are assumed. Table 4 shows the basic information about the cases.

Each case consists of five studies with the corresponding hypothesized population correlation coefficient and the sample sizes. For example, case 3 has the population correlation coefficient $\rho = -0.3$ and the five studies have the sample sizes 4,4,4,4 and 4, for each.

| Table 4: Four Cases used for Simulation | | |
|---|---|---|
| Case | Population correlation | Sample sizes |
| 1 | $\rho = 0.2$ | $n_1 = n_2 = n_3 = n_4 = n_5 = 4$ |
| 2 | $\rho = 0.2$ | $n_1 = n_2 = n_3 = n_4 = n_5 = 100$ |
| 3 | $\rho = -0.3$ | $n_1 = n_2 = n_3 = n_4 = n_5 = 4$ |
| 4 | $\rho = -0.3$ | $n_1 = n_2 = n_3 = n_4 = n_5 = 100$ |
| 5 | $\rho = 0.6$ | $n_1 = n_2 = n_3 = n_4 = n_5 = 4$ |
| 6 | $\rho = 0.6$ | $n_1 = n_2 = n_3 = n_4 = n_5 = 100$ |
| 7 | $\rho = -0.7$ | $n_1 = n_2 = n_3 = n_4 = n_5 = 4$ |
| 8 | $\rho = -0.7$ | $n_1 = n_2 = n_3 = n_4 = n_5 = 100$ |

In each case, the asymptotic pooled estimator and the MLEs (one from the exact likelihood and the other from a *B*-spline approximation) are calculated 100 times and the corresponding MSEs are also calculated.

Table 5 shows the MSEs of the asymptotic pooled estimator and MLEs for each case. Consistently, the MLEs show better performance than the asymptotic pooled estimator. The MLEs have practically the same MSEs for all but cases with very small sample sizes and relatively big $\rho$s in magnitude (case 5 and case 7), but even these cases do not show considerable differences between two methods. Case 2, case 4, case 6 and case 8 show that all the estimators work well when the sample sizes are large enough. Case 1, case 3, case 5 and case 7 show a very interesting trend when the sample sizes are small - the MLEs are much better than the asymptotic pooled estimator when the magnitude of $\rho$ is relatively small (case 1 and case 3) but the merit of getting the MLEs is diminishing as the magnitude of $\rho$ is increasing even though the MLEs are still better than the asymptotic estimator.

| Table 5: The MSEs from the Asymptotic, Exact Likelihood and B-spline Approximation | | | |
|---|---|---|---|
| Case | Asymptotic Pooled Estimate | MLE (Exact Likelihood) | MLE (B-spline Approximation) |
| 1 | 0.127 | 0.083 | 0.082 |
| 2 | 0.002 | 0.002 | 0.002 |

| 3 | 0.088 | 0.058 | 0.059 |
| 4 | 0.002 | 0.002 | 0.002 |
| 5 | 0.060 | 0.044 | 0.048 |
| 6 | 0.001 | 0.001 | 0.001 |
| 7 | 0.030 | 0.027 | 0.029 |
| 8 | 0.001 | 0.001 | 0.001 |

## 7. Discussion

In meta-analysis, asymptotic normal approximation approaches (pooled or weighted averages) have been used to combine information about a population correlation coefficient from many independent studies. But these approaches are questionable when we do not have large enough sample sizes. (See the result in section 6) The likelihood approach using *B*-spline approximation to calculate the likelihood is used in this paper.

As we can see in the previous sections, a *B*-spline generally gives a close approximation to the log-likelihood function. When the MLE of the combined likelihood (or log-likelihood) is not far from the origin, say between -0.8 and 0.8, a *B*-spline fits the exact likelihood function nearly perfectly (See Figure 2). From Table 3, we can also observe that *B*-splines detect the right location of the MLE and the likelihood interval from the exact combined likelihood, which are very important for inference. Furthermore, just 2 or 3 inner knots, which correspond to degrees of freedom 5 or 6 when we use a cubic *B*-spline in the computer program *R*, are used for approximation in the previous sections. Consequently, it is possible to obtain a parsimonious model from the exact likelihood function.

But there is a limitation when applying a *B*-spline approximation. When the MLE of $\rho$ is close to 1 in magnitude, the B-spline approximation fits less well, which is a general flaw of spline approximation when dealt with values close to the boundaries of domain of the approximated function. Fortunately, the vitamin C data do not suffer from this boundary issue, so the results are reliable.

I and Dr. Leon Gleser published the paper about combining multi-dimensional correlation matrices with a numerical integration methods, which do not cover the 2-dimensional case covered in this article due to mathematical derivation working for 3 or higher dimension. [13] This paper works as a complementary work for our paper and suggests a possible way to resolve the issue of 'curse of dimensionality' in the previous research by using *B*-spline approximation for future research.